# Printable Nanoscopic Metamaterial Absorbers and Images with Diffraction-Limited Resolution

*Patrizia Richner, Hadi Eghlidi\*, Stephan J. P. Kress, Martin Schmid, David J. Norris, Dimos Poulikakos\**

P. Richner, Dr. H. Eghlidi, M. Schmid, Prof. D. Poulikakos

Laboratory for Thermodynamics in Emerging Technologies, ETH Zurich, Sonneggstrasse 3, 8092 Zurich, Switzerland

E-mail: eghlidim@ethz.ch; dpoulikakos@ethz.ch

S. J. P. Kress, Prof. D. J. Norris

Optical Materials Engineering Laboratory, ETH Zurich, Leonhardstrasse 21, 8092 Zurich, Switzerland

Abstract: The fabrication of functional metamaterials with extreme feature resolution finds a host of applications such as the broad area of surface/light interaction. Non-planar features of such structures can significantly enhance their performance and tunability, but their facile generation remains a challenge. Here, we show that carefully designed out-of-plane nanopillars made of metal-dielectric composites integrated in a metal-dielectric-nanocomposite configuration, can absorb broadband light very effectively. We further demonstrate that electrohydrodynamic printing in a rapid nanodripping mode, is able to generate precise out-of-plane forests of such composite nanopillars with deposition resolutions at the diffraction limit on flat and non-flat, substrates. The nanocomposite nature of the printed material allows the fine-tuning of the overall visible light absorption from complete absorption to complete reflection by simply tuning the pillar height. Almost perfect absorption (~95%) over the entire visible spectrum is achieved by a nanopillar forest covering only 6% of the printed area. Adjusting the height of individual pillar groups by design, we demonstrate on-demand control of the gray scale of a micrograph with a spatial resolution of 400 nm. These results constitute a significant step forward in ultra-high resolution facile fabrication of out-of-plane nanostructures, important to a broad palette of light design applications.

**Introduction**

The rapid growth of novel metamaterials and devices serves as a driver for the continuous development of related micro- and nanofabrication technologies. Due to their simplicity and additive nature, printing processes are particularly attractive for such applications and are, as a result, presented with a broad palette of exciting opportunities and challenges. Three-dimensional microprinting enables the fabrication of devices of sizes down to a few tens of microns.[1-5] With respect to the resolution of planar printed pictures, the diffraction-limit (accessible with comparatively facile far-field optics) is being reached with the help of plasmonic structures.[6-9] It has been shown that intelligent structuring of the top metal layer of a metal-insulator-metal (MIM) configuration enables the fine-tuning of the absorption spectrum, paving the way toward reproduction of color images at the diffraction limit. However, previous fabrication approaches of such structures have always included expensive and difficult to upscale steps, such as electron-beam lithography or focused ion beam.

We show that metal-dielectric composite nanopillars in an MIM configuration can act as extraordinarily broadband absorbers of light within a diffraction-limited pixel. Their level of absorption can be fine-tuned by varying either the height or diameter of the pillars, their spacing or their material composition. To realize composite nanopillars, we use an electrohydrodynamic printing method termed NanoDrip printing[10]. Previously, with this technique, we were able to print colloids of various materials ranging from quantum dots[11] to metallic nanocrystal particles[12-13] in plane with feature sizes down to a few tens of nanometers. The technology proved to be very versatile in respect to nanoparticle size and material. In this paper we advance the unique capabilities of this methodology to print out-of-plane nanostructures on non-planar, rigid or soft substrates. We demonstrate printing of arrays of composite nanopillars with height, diameter and spacing specificity, which complements earlier works on pillar arrays based on silicon substrates where the pillar height is constant over the area of one sample.[14-15] We show in parallel that the material composition and nanofabrication precision of the printed out-of-plane nanopillar patterns enable the tuning of their overall absorption, a necessary requirement to achieve true submicron gray-scale resolution. Finally, we demonstrate how with this technique gray scale images at resolutions close to the diffraction limit can be readily fabricated in an open atmosphere.

**Results and Discussion**

The details of the printing process can be seen in **Figure** 1a. A gold-coated thermally drawn and tapered glass pipette with an opening of 1000-1500 nm is filled with a nanoparticle-laden dispersion (ink) and connected to a high voltage source. The substrate is placed on an electrically grounded piezo-stage and brought within 5 µm of the pipette tip. By applying a DC voltage of 125-175 V during the time period required for printing each nanostructure, attoliter-sized ink droplets (50-100 nm diameter) of precisely controllable volume are electrohydrodynamically pulled out of the apex of the stable meniscus at the end of the pipette. The droplets land on the substrate and their solvent evaporates before the arrival of the next droplet, leaving behind only its nanoparticle content confined in a substrate region within the initial diameter of the droplet. While the first few droplets printed on one spot land on an area of about twice the droplet diameter, they rapidly build an out-of-plane structure which acts as a lightning rod within the applied electric field and the following droplets land precisely on top of the so grown structure. In this fashion an out-of-plane pillar with the diameter equal to the droplet diameter[10] is created. Details of the electrohydrodynamic NanoDrip printing (EHD NanoDrip printing) process can be found elsewhere.[10, 16] While the technology is currently in its developing stage, first attempts are being made to enhance the serial single-nozzle output to a much faster multi-nozzle process. This will reduced the processing time, which is currently proportional to the printed area. The diameter of a pillar can be flexibly varied from 50 to 150 nm by changing the pipette opening diameter or the applied voltage (Figure 1b-d). The height of each printed pillar is proportional to the number of ejected droplets and is hence defined by the duration of the applied electrical pulse (Figure 1 e-g). Therefore, the height, position and diameter of each pillar can be controlled on demand using our home-built electrical and positioning control units.

In this work we were able to print reproducibly out-of-plane pillars with a diameter of 120 nm and heights in the range of 200-1000 nm in a hexagonal arrangement with a pillar-to-pillar spacing of 500 nm. This corresponds to a maximum aspect ratio of the pillars of more than 8. While for single pillars aspect ratios of up to 17 have been shown in an earlier publication[10], when printed in a dense array, neighboring pillars bend towards the applied electric field due to their soft nature[16] and it is not possible to obtain a well-defined array of such vertical pillars. The minimum pillar-to-pillar spacing is therefore limited by the maximal height of the neighboring pillars. With shorter pillars however, it is possible to print arrays with a smaller pillar-to-pillar spacing (see also Figure S5, Supporting Information). In Figure 1h a scanning electron micrograph (SEM) of pillar arrays is displayed at a 30° viewing angle. Our bottom-up approach allows the precise deposition of such pillar arrays also on non-planar substrates. We demonstrate this by printing pillars near and exactly on top of a template-stripped *silver* wedge[11,17] with a wedge angle of 70 degrees (Figure 1 i,j). This capability enables NanoDrip printing to overcome a major drawback of mask-dependent technologies such as nano-imprint lithography, which require a flat substrate. Furthermore, our approach can be used to print on soft substrates (see Figure S4, Supporting Information).

No annealing post-treatment is applied to the printed pillars, such that they consist of a porous scaffold of single nanoparticles. In **Figure 2a** a pillar array has been scratched with tweezers to expose their internal structure. The close-up SEM micrograph clearly shows the porosity both along the sides of the pillar as well as in the cross section, as can be seen in the cut-off base in the upper right. The arrangement of the gold nanoparticles during the build-up of a pillar is shown in the SEM micrograph in Figure 2b, where the pillar printing was terminated after the ejection of a few droplets. The semi-ordered arrangement of the gold nanoparticles after deposition is discernable.

In this paper we employ three different inks, which are composed of 5 nm spherical gold nanoparticles dispersed in tetradecane and coated with either octanethiol, decanethiol or dodecanethiol monolayers. While the printing behavior itself is not influenced by the different length of the ligands, the nanoparticle-to-nanoparticle distance in the printed gold-air nanocomposite scaffold is a linear function of the alkyl chain length.[18] This distance in turn influences the gold-to-air ratio in the printed structures. To model the electrical properties of printed structures, needed for the quantification of their interaction with light, we use the Bruggeman's effective medium theory[19] which considers the printed gold-air nanocomposite as a homogenuous medium with an effective dielectric function given by

$$\varepsilon = \frac{1}{4}\left( (3FF-1)\varepsilon_{gold} + (2-3FF)\varepsilon_{air} \pm \sqrt{\left((3FF-1)\varepsilon_{gold} + (2-3FF)\varepsilon_{air}\right)^2 + 8\varepsilon_{gold}\varepsilon_{air}} \right) \quad (1)$$

where FF is the volumetric filling fraction of gold and $\varepsilon_{air}$ and $\varepsilon_{gold}$ are the permittivities of air and gold, respectively. The sign is chosen such that the imaginary part of the dielectric function is positive. The simulation software LUMERICAL[20] which is based on a finite-difference time-domain (FDTD) method is used to numerically model the fabricated structures. Both in experiment and simulations, the pillars are placed on a 60 nm thick layer of $SiO_2$ deposited on an optically thick gold layer (inset **Figure** 3b). It was previously shown that thin layers of gold-dielectric nanocomposites are very good absorbers of visible light.[21] This is due to an increase in the imaginary part of the effective permittivity of nanocomposites, which is responsible for the absorption, as compared to the ones of pure metal and dielectric.[19] Unlike most of the previously reported metal-insulator-metal absorbers,[22-23] having the nanopillars on top of a reflector enables absorption of the impinging light at the nanocomposite material both at direct incidence and also after reflection by the gold layer. This increases the total absorption with the same surface coverage of the pillars, and therefore provides a larger range of absorption tunability. While for the dielectric function of the gold back-reflector the data from Johnson and Christy[24] were used, the permittivity of the pillars is calculated from equation (1). The gold nanoparticles in a printed pillar are loosely and randomly packed. The packing density of such particles has been shown theoretically to be 55% for touching spheres.[25] Here the outer shell of the spheres consists of the alkyl derivatives, the thickness of which we know from interparticle distance measurements of gold nanoparticles with thiol-capped alkyl derivatives.[18] The volume of the outer shell is hence subtracted from the nanoparticle FF. The FF values used for the simulations for all three cases studied in this work can be found in Table 1. Furthermore, according to Kreibig and Fragstein,[26] the dielectric function of particles that are small compared to the mean free path of the conduction electrons, becomes a function of the particle diameter. The dielectric function of bulk gold is hence adapted for nanoparticles with a diameter of 5 nm (see Figure S2, Supporting Information).

Figure 3a shows the measured absorption spectra of pillar arrays printed with the decanethiol-capped ink. Increasing the pillar height increases the total absorption. 95% absorption over the entire visible spectrum and even into the near infrared is achieved with a pillar height of 560 nm. An extraordinary point here is that this nearly perfect absorption over the entire surface is achieved with a pillar array that covers only 6% of the surface (top view). As indicated by the dashed line, the gold back reflector has an intrinsically high absorption between 450 and 500 nm, yielding the familiar yellow color of the noble metal. Simulations of arrays with the same height as in Figure 3a are depicted in Figure 3b.

Figure 3c shows the measured absorption spectra for absorbers printed with inks with three different ligands, i.e. octanethiol, decanethiol and dodecanethiol. The calculated FF values for each case are reported in Table 1. While the absorption over the visible range is comparable in all three cases, the drop-off wavelength is a function of the filling fraction. The gold-air nanocomposite filling fraction influences the wavelength range where the effective electrical conductivity is substantial. The electrical conductivity is responsible for ohmic losses, hence the overall absorption band can be tuned using the FF value. The broad absorption spectrum is due to the fact that in all the three cases we have an FF close to the percolation limit (33%).[19] At the percolation threshold the imaginary part of the permittivity of a metal-dielectric nanocomposite, which is responsible for the ohmic losses in the nanocomposite, becomes very large over an infinitely broad spectral range.[19]

The simulated spectra in Figure 3d also show a similar trend as the measurements in terms of the absorption levels and drop-off wavelengths. However, some of the spectral feature details in the simulations are not observed in the measurements. The reason can be attributed to deviations from the ideal structures simulated, such as to the surface roughness and non-homogeneities of the printed structures, which are not accounted for in the simulations and also to the non-ideality of the Bruggeman's theory in approximating the effective permittivity of the printed structures (equation (1)). A more accurate method to model the optical properties of the nanocomposite structures can be based on full-wave simulation of the random arrangement of the nanoparticles in the nanocomposite.[27] However, this will be numerically very expensive.

Figure 3e shows the simulated absorption as a function of wavelength λ and pillar height. Here, the excellent tunability of the absorption spectrum achieved by only changing the pillar height is demonstrated. The gray scale range can be covered practically in its entirety by printing pillars with heights ranging from zero (perfect reflection, white) to 650 nm and above (nearly perfect absorption, black). In the current study, the absorption at lower wavelengths (below 550 nm) is maintained even for pillar height equal to zero. This is a result of intrinsic losses of bulk gold and when the pillar height is zero, absorption takes place entirely in the gold back reflector. To isolate the pure effect of pillar absorption if so desired, one can use silver instead of gold as the back reflector (see Figure S5, Supporting Information). A thin silver layer has negligible absorption and is often used as a perfect mirror in the visible range, yielding white color (perfect reflection) in the absence of pillars.

We calculate the total absorption of sunlight (standard light source) as a function of pillar height with the following equation:

$$A_{tot} = \frac{\int A(\lambda)I(\lambda)d\lambda}{\int I(\lambda)d\lambda} \tag{2}$$

where $A(\lambda)$ and $I(\lambda)$ are the absorption reported in Figure 3e and the solar irradiance for the standard AM1.5 (air mass) spectrum, respectively, the integration is performed over the visible spectrum, 400-700 nm. The perception of light intensity coming from a reflecting surface to the human eye, is proportional to the logarithm of the light reflected from the surface.[28] Therefore, the semilog plot of reflection (= 1 − absorption, where absorption is calculated from equation (2)) reported in Figure 3f represents the eye's perception of the brightness of the printed absorbers with different pillar heights. Changing the pillar height, we can tune this brightness practically across the entire gray scale range. The optical images of three arrays of printed pillars with various heights are displayed in the micrographs in the inset to Figure 3f. Three points corresponding to these three arrays are indicated on the curve in Figure 3f.

Next we used the decanethiol gold nanoparticle ink to investigate the effects of pillar diameter and pillar-to-pillar spacing. Graphs of the corresponding measurements and simulations can be seen in Figure 4. In Figure 3 we analyzed arrays with a pillar diameter of approximately 120 nm. In Figure 4 (a: experiments, b: FDTD simulations) we show the absorption spectra of three different pillar diameters, all with a pillar height of about 470 nm. The absorption increases with increasing pillar diameter. However, there is an upper limit to the pillar diameter, which can be printed reproducibly in an array with 500 nm pillar-to-pillar spacing: For diameters larger than 120 nm it is difficult to print regular arrays with a pillar height of 470 nm. The greater proximity of pillars to each other leads to a negative influence on the printing process, in that the droplets are diverted towards the already printed pillars, rendering the realization of a regular array prohibitively difficult.

In Figures 4c and d (experiments and simulations, respectively) we studied the influence of the pillar-to-pillar spacing on the absorption spectrum.. The pillar height in this case is 700 nm and the diameter 100 nm. The absorption decreases with increasing pillar-to-pillar spacing, which can be understood by considering that by increasing the spacing we deposit less of the nanocomposite absorbing material per area. In addition to being able to largely tune the absorption of each pillar by simply varying its height or diameter, another important advantage of the NanoDrip printing method lies in its ability to place pillars at small distances (tight pitch control). These properties enable printing gray scale images with sub-micron pixel size (spatial resolution) determined by the pillar-to-pillar distance, which in this work is 500 nm. The latter can be even further reduced for pillars smaller than 1000 nm in height (see also Figure S5 in Supporting Information with 400 nm pillar-to-pillar pitch).

**Figure 5** shows an exemplary picture printed with the method reported in this work. The 395x353 pixel sized picture of Charlie Chaplin (adapted from Charlie Chaplin TM © Bubbles Incorporated SA) is printed in 8 shades of gray ranging from dark to bright. Seven different pillar heights are used to produce the darker shades and white corresponds to no pillar printed. In order to adequately represent the pixels, the pillars are printed in a Cartesian grid with a pixel-to-pixel spacing of 500 nm, resulting in an image size of ca. 200x175 µm, corresponding to 50'000 pillars per inch and a resolution of out-of-plane printed pillars close to the diffraction limit. This makes the method presented here comparable to electron beam lithography (EBL) based methods,[6-7] featuring more than 2 orders of magnitude higher resolution in printing *out-of-plane* structures than a standard commercial inkjet printer for planar structures. The scanning electron micrographs in Figures 5c,d show that while the pillars look almost identical from the top (Figure 5c), they are designed to vary in height as desired (30° tilted view, Figure 5d) and therefore yield different shades of gray in an optical image, enabled by controlling the height of every single pillar. The supplementary Fig. S5 shows another printed gray-scale image with a resolution of 400 nm. For this image we used a silver back-reflector, which leads to a perfect reflection (white) in the absence of nanopillars.

**Conclusion**

In closing, we were able to directly print in an open atmosphere arrays of closely-spaced, out-of-plane nanopillars made of a gold-air composite. Adjusting the pillar heights in a metal-insulator-nanocomposite plasmonic configuration, we demonstrated tuning of the absorption performance of the printed arrays from total absorption (dark) to total reflection (white). Maximum average absorption of 95% over the entire visible spectrum (up to 800 nm in wavelength) was achieved by an array of nanopillars, which cover only 6% of the entire printed area. We showed, theoretically and experimentally, that the bandwidth and absorption level of the printed absorbers can be controlled independently by varying the metal filling fraction and pillar height, respectively. Furthermore, we demonstrated the printing of gray-scale micrographs with ultra-small pixel sizes (500 and 400 nm) by adjusting the height of each pillar in an array. Other main advantages of our method include the flexibility to print on non-flat substrates and its independence of any mask and of expensive clean-room steps. The capability to realize the features presented here is relevant to a host of applications, such as the generation of individualized security or proof of authenticity features invisible to the naked eye, and optical data storage.

**Methods**

*Ink Preparation:* The synthesis of the gold nanoparticles follows a published recipe.[29-31] It results in nanocrystals of approximately 5 nm in size (see TEM image in Figure S1, Supporting Information). The ligand exchange of the originally dodecanethiol-capped nanoparticles to the desired alkylthiolate capping ligand (octanethiol, decanethiol) follows an earlier published recipe.[32-33] The concentration of the final ink is approximately 13 mg Au per ml tetradecane.

Absorption spectra and details of the preparation process can be found in the Supporting Information.

*Printing:* A clean silicon wafer is Ar-sputtered for 60 s; first 100 nm Au (Plassys E-beam) and then 60 nm $SiO_2$ (Oxford Instruments PECVD 80Plus) is deposited. Details of the printing mechanism can be found elsewhere;[10] the piezo stage (Mad City Labs XYZ Nano-Drive) motion was controlled by an in-house built control unit. The pulse lengths for the pillar generation were varied from 500 ms to 2500 ms to print pillars of increasing height. The pillars for the absorption measurements were printed in a hexagonal array with a 500 nm pitch.

In order to represent the picture in Figure 3 to scale, a Cartesian grid was employed. The 7 shades of gray in the original picture were represented by 7 pillar heights (250-1750 ms pulse length per pillar), the lightest gray corresponding to the lowest and the darkest to the highest pillar.

*Characterization*: Absorption measurements are performed using a home-built inverted microscope equipped with an air objective (numerical aperture 0.75) for exciting the sample and collecting the reflected and scattered light. The absorption spectra are obtained in the range from 400 to 800 nm with an Acton SP2500 spectrograph equipped with an eXcelon detector from Princeton Instruments. A xenon lamp was used as a light source. Several measurements were taken over the course of a few days and after long exposure of the samples to the xenon lamp to ensure they did not degrade.

The absorption was calculated according to the following equation:

$$A = \frac{L-R}{L-DC} \tag{3}$$

where A is the absorption, L is the spectrum of the lamp measured with a silver mirror, R is the spectrum of light reflected and scattered by the pillar arrays, and DC is the intrinsic spectral dark count of the spectrometer.SEM micrographs were taken with a Zeiss ULTRA plus scanning electron microscope after sputter-coating a thin Au/Pd layer onto the samples to reduce charging of the non-conductive sample, unless otherwise stated.

**Supporting Information**

Supporting Information. Ink preparation, FDTD simulation Parameters, TEM images, absorption of unstructured layer (PDF).

The authors declare the following competing financial interest(s): One of the authors (D.P.) is involved in a startup company that is attempting to commercialize the printing process used in our manuscript.

Acknowledgements

We gratefully acknowledge assistance from V. Holmberg. The research leading to these results has received funding from the Swiss National Science Foundation under Grant 146180. We also gratefully acknowledge funding from the European Research Council under the European Union's Seventh Framework Programme (FP/2007-2013) / ERC Grant Agreement Nr. 339905 (QuaDoPS Advanced Grant).

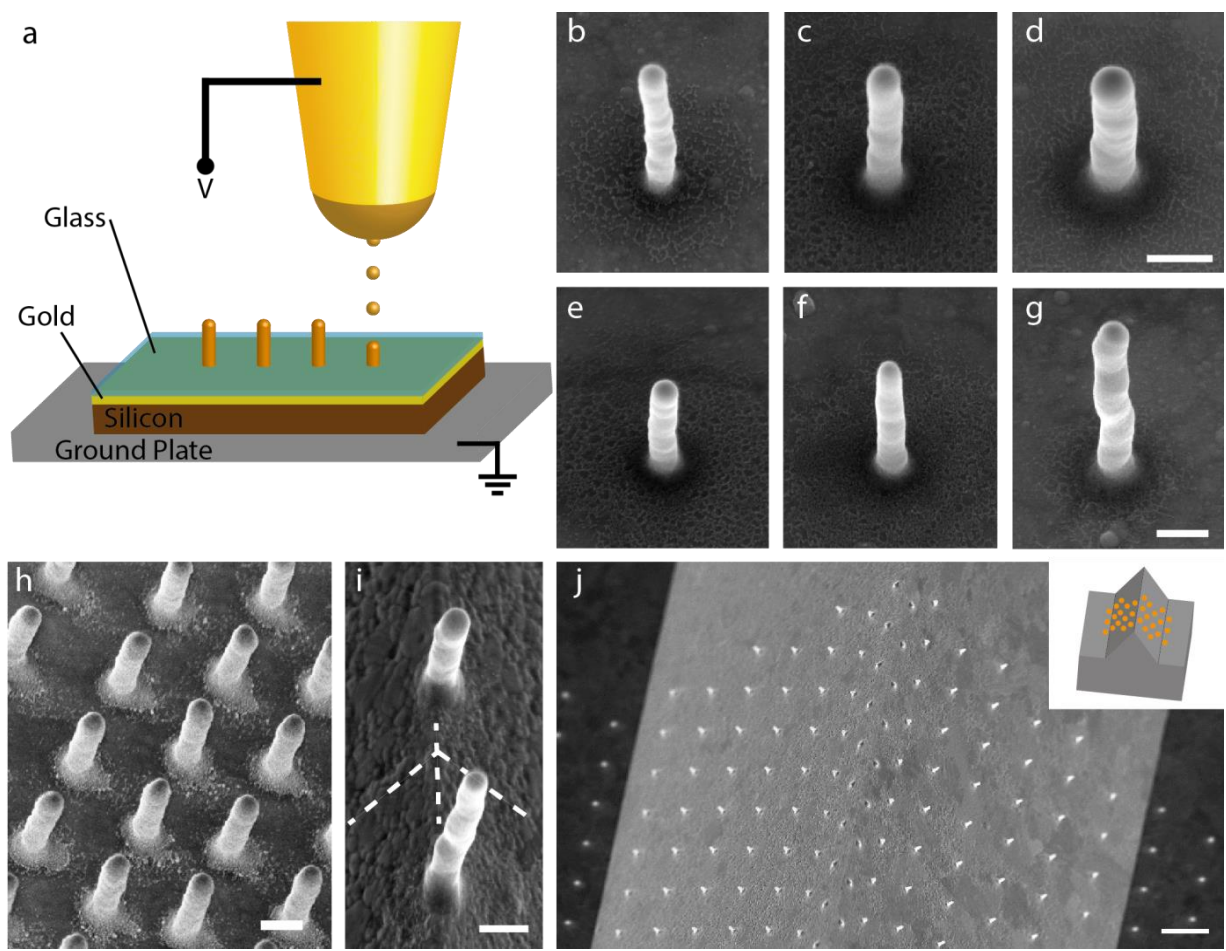

**Figure 1.** a) Schematic of the printing mechanism. The thickness of the gold and the SiO$_2$ layers are 100 nm and 60 nm, respectively, and the pillar heights can be controllably varied. Scanning electron micrographs: b)-d) pillars of different diameter, controlled by the nozzle size and the applied voltage. e)-g) pillars of different height, controlled by the time duration of the applied voltage pulse; h) hexagonal array of pillars with a 500 nm pillar-to-pillar spacing; i) pillars printed precisely on top of the edge of a template-stripped silver wedge with 70.5° top angle; j) pillar array printed on the surface of a wedge, wedge height is 15 µm, inset: schematic of 3D substrate with pillars. Scale bars correspond to 200 nm (b-i) and 2 µm (j).

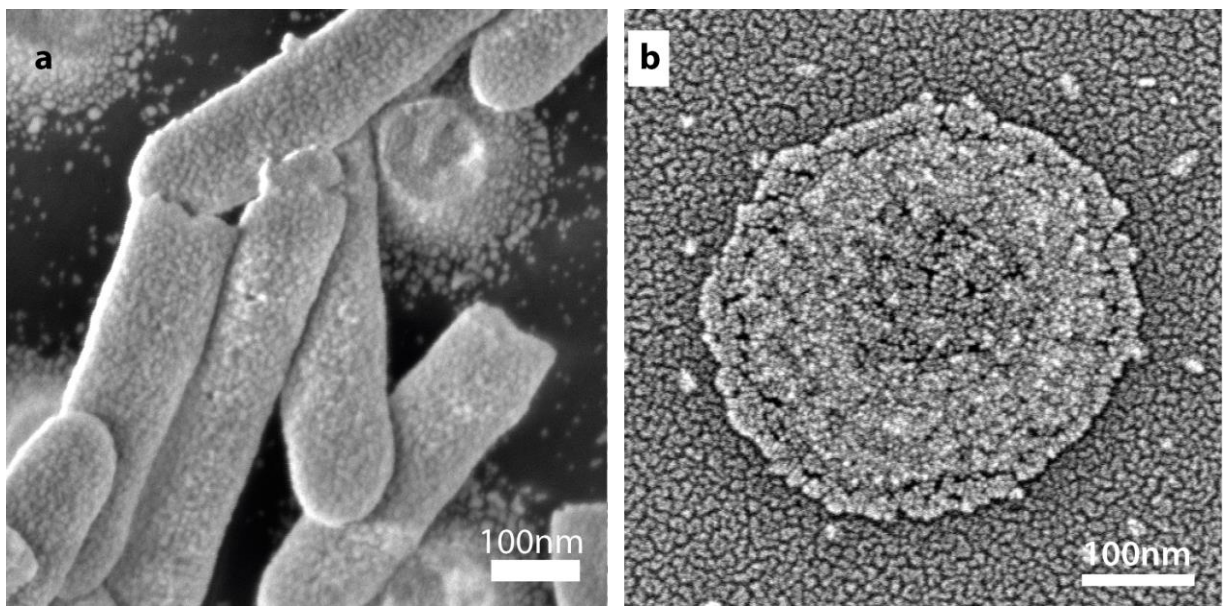

**Figure 2.** a) Close-up SEM micrograph of pillars scratched by tweezers. The porous structure at the surface of the pillar as well as in the middle of the broken base in the upper right corner is discernable. This sample was not sputter-coated to fully assess the original porosity of the structures; b) several ejected droplets at the same location. The arrangement and layering of the gold nanoparticles is clearly visible. This sample was sputter-coated with a few nm of Pd/Au for better contrast.

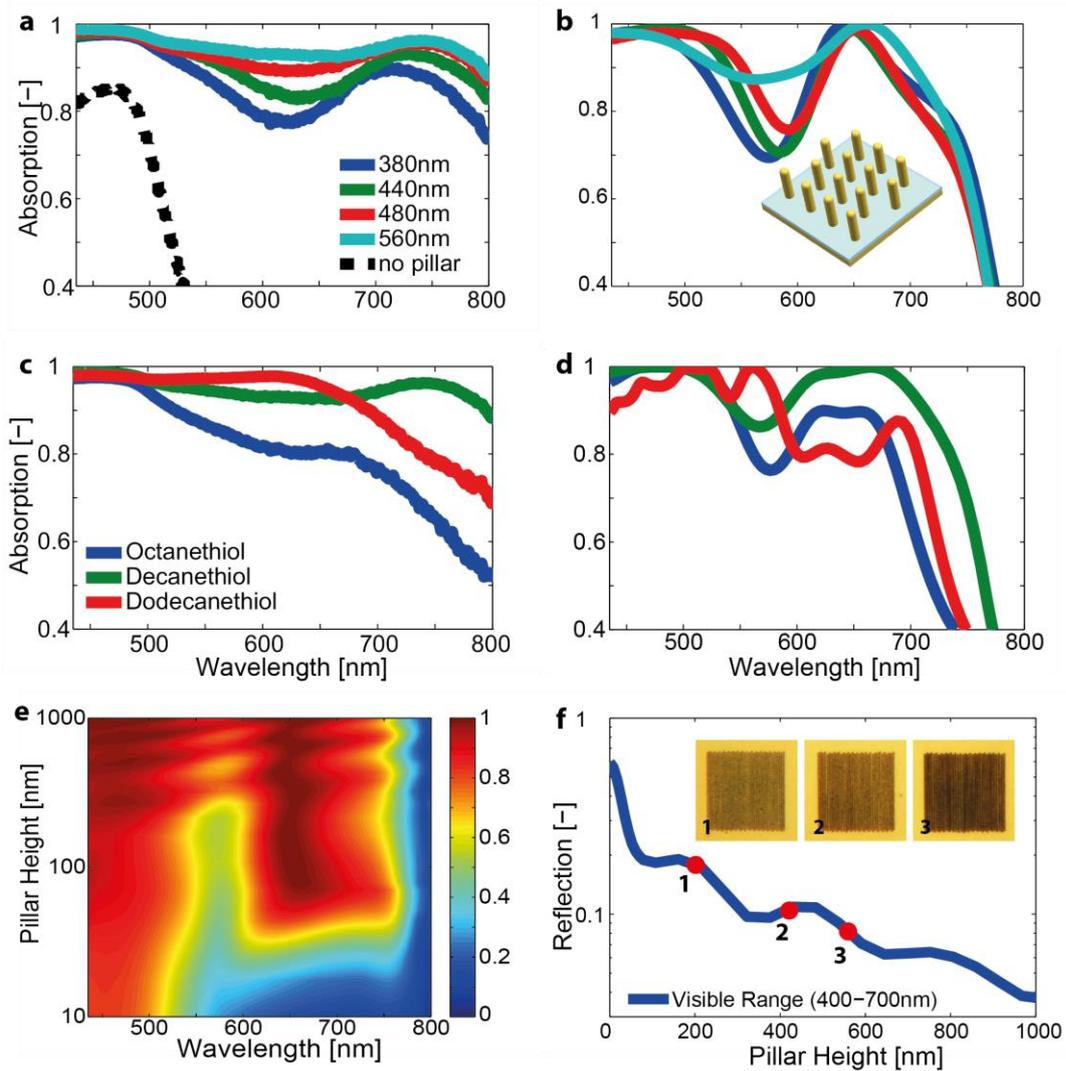

**Figure 3.** Measurements (a) and simulations (b) of absorber arrays with different pillar heights. Dashed line in a) is the absorption spectrum of the gold-glass substrate before printing. Measurements (c) and simulations (d) of absorber arrays with different inks. The pillar height in all three cases is approximately 600 nm; e) color plot of absorption spectrum with various pillar heights; f) Total reflection of the solar spectrum (AM1.5) as a function of pillar height, based on the simulations displayed in (e). The inset shows optical images of three absorber arrays with pillar heights of 200, 420 and 560 nm. The three points on the plot correspond to the printed arrays.

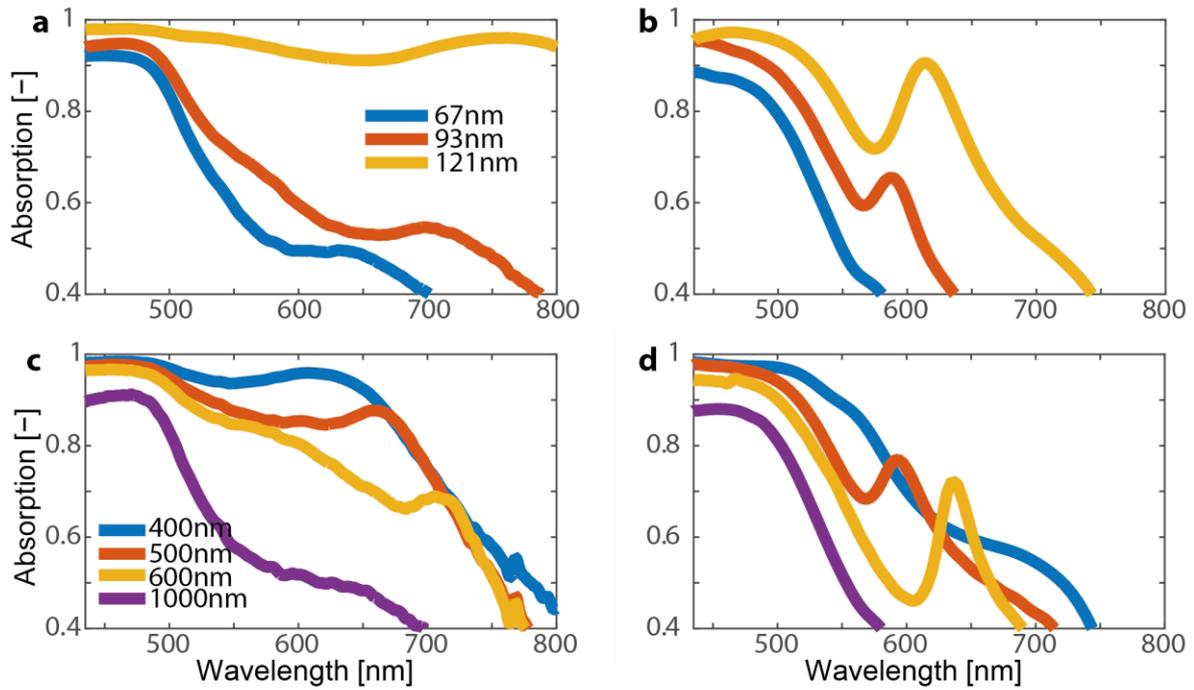

**Figure 4.** The effects of pillar diameter (a),(b) and pillar-to-pillar spacing (c),(d) on absorption. (a) and (c) are experimental and (b) and (d) are simulation results. The parameters for the different plots are given in Table 2.

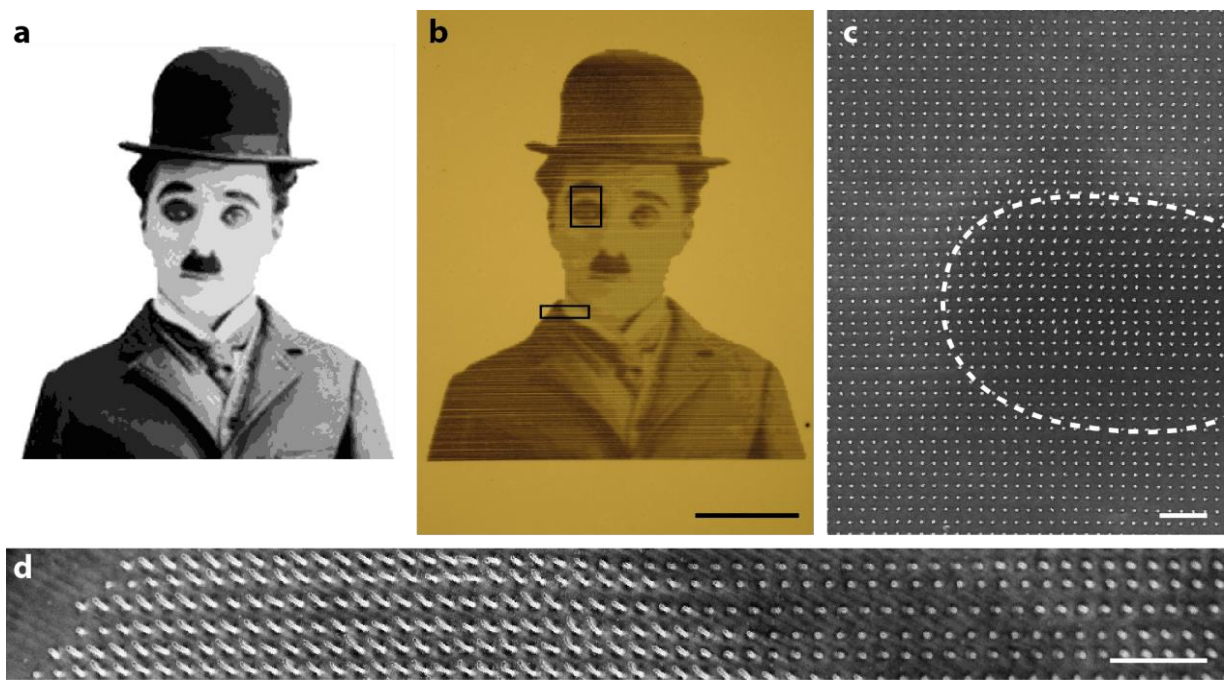

**Figure 5.** a) black-and-white photograph of Charlie Chaplin, adapted from Charlie Chaplin TM © Bubbles Incorporated SA; b) brightfield optical micrograph of a printed reproduction with a 500 nm pixel size, the rectangles represent the regions from which images in c) and d) are acquired; c) scanning electron micrograph of the area around the eye. The perimeter of the eye is indicated with the dashed line. While the pillar diameter remains the same for darker and brighter pixels, the increased height of the pillar leads to decreased brightness of the pixel; d) a 30° tilted view scanning electron micrograph of the collar area. The height difference of bright and dark areas is clearly discernible. Scale bars correspond to 40 µm (b), and 2 µm (c,d).

**Table 1.** Interparticle distance and volume filling fraction of different thiol-coated nanocrystals.

| Surfactant | Interparticle distance[18] [nm] | Volume filling fraction (FF) of random loose packed ligand-coated gold nanospheres |
|---|---|---|
| Octanethiol | 1.45 [a] | 25.6% |
| Decanethiol | 1.65 | 23.4% |
| Dodecanethiol | 1.85 | 21.4% |

[a] value interpolated from [18]

**Table 2.** Parameters used for the plots in Figure 3

|  | (a) | (b) | (c) | (d) |
|---|---|---|---|---|
| Pillar-to pillar spacing | 500 nm | 500 nm | Variable | Variable |
| Pillar height | 470 nm | 470 nm | 700 nm | 700 nm |
| Pillar diameter | Variable | Variable | 100 nm | 100 nm |
| Surfactant | Decanethiol | | | |
| Data source | Measurement | Simulation | Measurement | Simulation |

**Table of Contents Graphic**

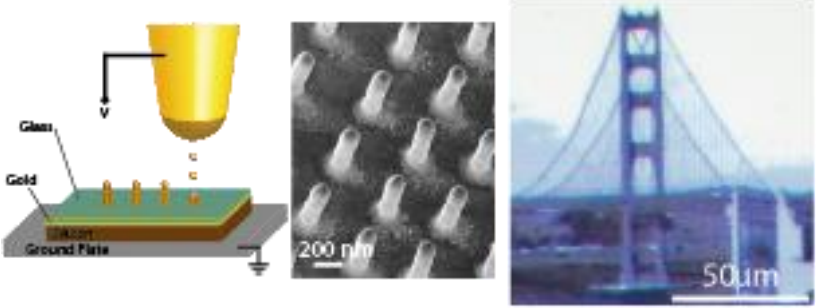

# Supporting Information

# Printable Nanoscopic Metamaterial Absorbers and Images with Diffraction-Limited Resolution

*Patrizia Richner, Hadi Eghlidi\*, Stephan J.P. Kress, Martin Schmid, David J. Norris, Dimos Poulikakos\**

P. Richner, Dr. H. Eghlidi, M. Schmid, Prof. D. Poulikakos
Laboratory for Thermodynamics in Emerging Technologies, ETH Zurich, Sonneggstrasse 3, 8092 Zurich, Switzerland
Email: eghlidim@ethz.ch, dpoulikakos@ethz.ch

S. J.P. Kress, Prof. D. J. Norris
Optical Materials Engineering Laboratory, ETH Zurich, Leonhardstrasse 21, 8092 Zurich, Switzerland

## Ink Preparation

**Materials.** Chlorotriphenylphosphine gold ((Ph3P)AuCl, 99.9% trace metal basis), ethanol (EtOH, 99.8%), decanethiol (DT, 96%), dodecanethiol (DDT, 98%), hexane (99%), octanethiol (OT, 98.5%), tetradecane (TD, 99%), and toluene (99.8%) were purchased from Sigma. Tert-butylamine borane (TBAB, >97%) was obtained from Strem Chemicals. All the chemicals were used as received without further purification unless otherwise stated.

**Synthesis of spherical gold nanoparticles (SGNP) with dodecanethiol (DDT) ligands.** The synthesis follows a published recipe[1-3]. Initially, 1 mmol of CTPPh-Au is dissolved in 100 mL toluene in a covered 200 mL beaker and heated under vigorous stirring to 100°C in an ambient atmosphere. Next, 2 mL of DDT was swiftly added, followed by the rapid injection of a solution of 8 mL toluene containing 20 mmol of the reducing agent TBAB, which led to the formation of the nanocrystals (rapid darkening from red to black of the solution). After keeping the solution at 100°C for 60 min, the dispersion was cooled down to room temperature while continuously stirring. To separate out larger agglomerates formed during the synthesis, the as-prepared solution was centrifuged and the supernatant retained. Next, the particles were precipitated by adding ethanol to the toluene solution (2:1) and subsequent centrifugation, discarding the clear supernatant and redispersing the nanocrystals in a total of 20 mL toluene. This precipitation and redispersion step was repeated and the nanocrystals were stored in the dark at ambient conditions until further use. The recipe resulted in nanocrystals of approximately 5 nm in size (see Figure S1a, Supporting Information) at a concentration of 10 mg/ml Au in toluene with the predominant amount of thiols bound to the surface of the nanocrystals.

**Ligand exchange of DDT-capped SGNP with other alkylthiols.** For the ligand exchange of the original DDT-capped SGNP we followed earlier published recipes[4-5]. To 5 mL of the stock solution (described in previous section), we added 1 mL of the desired thiol (octanethiol, decanethiol, dodecanethiol). This large excess along with sufficient time of 96h[5] ensured a nearly complete exchange to the new ligand while maintaining the particle size (see TEM images). The exchange was terminated by precipitating the nanoparticles using ethanol (1:1) and centrifugation. Discarding the supernatant, the precipitate containing the nanoparticles was redispersed in 2 mL of hexane. This precipitation-redispersion cycle using ethanol/hexane was repeated. Next, the nanocrystals were transferred to tetradecane by selective overnight evaporation of the hexane at room temperature in an open, wide-neck 20 mL scintillation vial. After the transfer, the optical density of the dispersion was adjusted (optical density of 50 at 450 nm for a 10 mm cuvette) using a spectrophotometer (Varian Cary 50) and a 1 mm quartz cuvette (Hellma Analytics). The resultant inks had approximately 25 mg Au per mL of tetradecane and were stored at ambient conditions in the dark until used for the printing.

**Preparation for printing.** For printing, the inks were further diluted with tetradecane to yield an optical density of 5. While this limits the material deposition speed, it ensures a stable printing process over the course of several hours without clogging at the nozzle. During the course of the experiments the stability and quality of all inks employed in this work has not decreased.

# FDTD Simulations Parameters

The dielectric function of a bulk metal can be decomposed into a bound- and a free-electron term:

$$\varepsilon_{bulk} = \varepsilon_{bound-electron} + 1 - \frac{\omega_p^2}{\omega^2 + i\omega\gamma_{bulk}} \qquad (1)$$

where $\omega_p$ is the bulk plasma frequency ($1.3 \cdot 10^{16}\ s^{-1}$) and $\gamma_{bulk}$ ($1.64 \cdot 10^{14} s^{-1}$) is the damping constant in the Drude model. $\varepsilon_{bound-electron}$ is obtained by subtracting the free electron part from $\varepsilon_{bulk}$.[6]

For particles which are small compared to the mean free path of the conduction electrons, the dielectric function differs from the bulk values. In that case collisions of the electrons with the particle surface gain importance as an additional damping factor. Kreibig and Fragstein[7] show that for small particles the effective mean free path becomes a function of the particle radius and the damping constant can be extended as follows:

$$\gamma(R) = \gamma_{bulk} + C\frac{v_F}{R} \qquad (2)$$

where $v_F$ is the Fermi velocity ($14.1 \cdot 10^{14} nms^{-1}$), R the nanoparticle radius and C a constant accounting for details of the scattering process. C=1 for isotropic scattering[6] was chosen for this work.

With the above formulas the dielectric function for the gold nanoparticles could be obtained. In order to also adequately represent the porous structure of the pillars, we used the Bruggeman effective medium theory.[8] Both the gold nanoparticles and the surrounding dielectric are treated equally in the mixture. The composition is defined by a gold volume fill factor f. The dielectric function of the effective medium for the pillars is then calculated as follows:

$$\varepsilon = \frac{1}{4}\left((3f-1)\varepsilon_{Au} + (2-3f)\varepsilon_{air} \pm \sqrt{((3f-1)\varepsilon_{Au} + (2-3f)\varepsilon_{air})^2 + 8\varepsilon_{Au}\varepsilon_{air}}\right) \qquad (3)$$

The sign in the above equation is chosen such that the imaginary part of the dielectric function is positive.

The interparticle distance of gold nanoparticles with thiol-capped alkyl derivatives of different lengths was measured by Wan et.al.[9] We assume here the length of the alkyl derivative to be half of the measured interparticle distance when packed in a pillar. The packing density of monodisperse, randomly packed spheres has been shown[10] to be 55% for random loose packing. In order to account for the volume taken up by the surfactant one needs to subtract the volume fraction filled by the spherical shell of the alkyl derivatives, the thickness of which is known from the interparticle distance measurements. The volume fraction and hence the fill factor of gold then reduces to less than half of the value for touching solid spheres.

According to Kreibig and Fragstein[7] the dielectric function of particles small compared to the mean free path of the conduction electrons, becomes a function of the particle diameter.

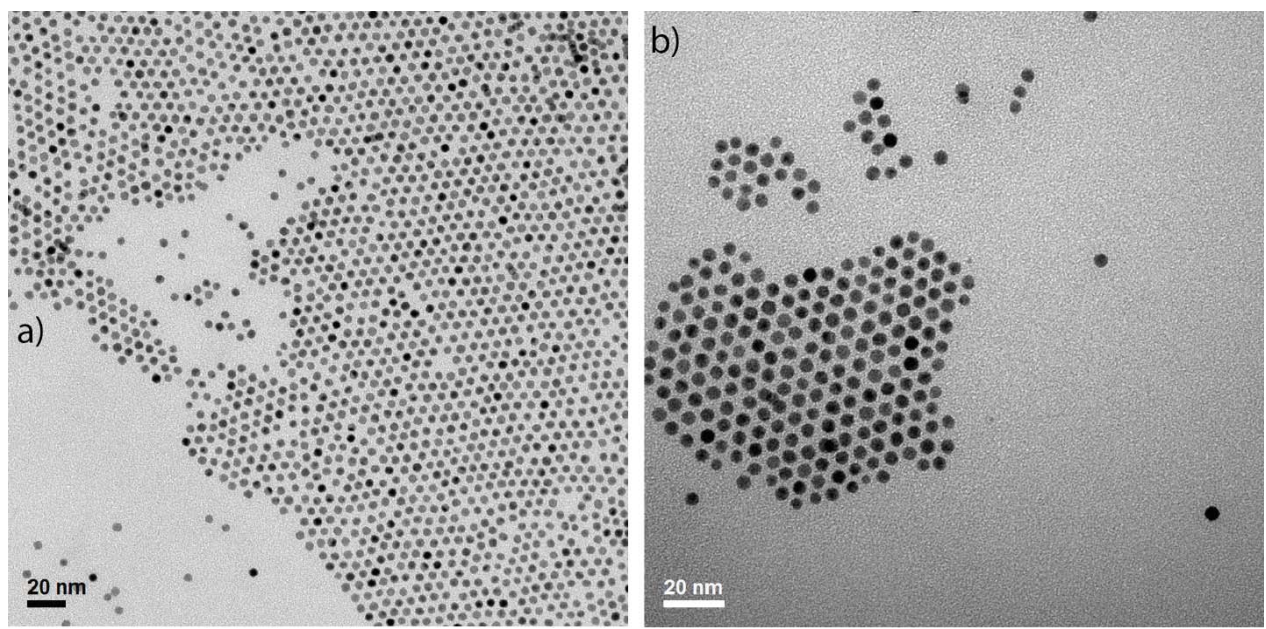

**Figure S1**. Transmission electron microscopy (TEM) image before ligand exchange (a) and after ligand exchange (decanethiol) (b)

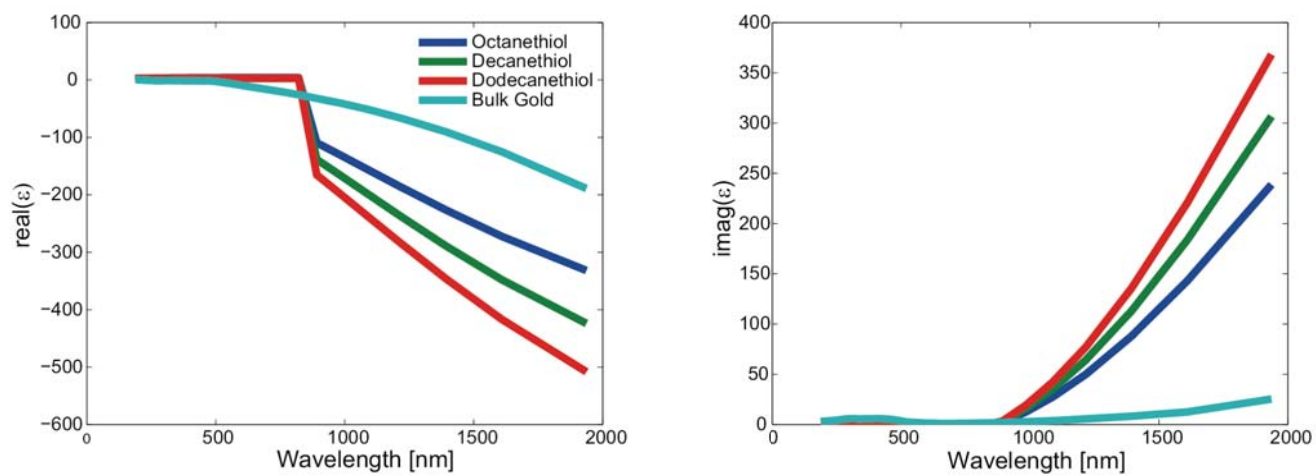

**Figure S2.** Dielectric function for all three ligands used in the simulations as well as bulk gold results as taken from Johnson and Christy[11]. The left panel shows the real part of the dielectric function, the right panel the imaginary part.

S-5

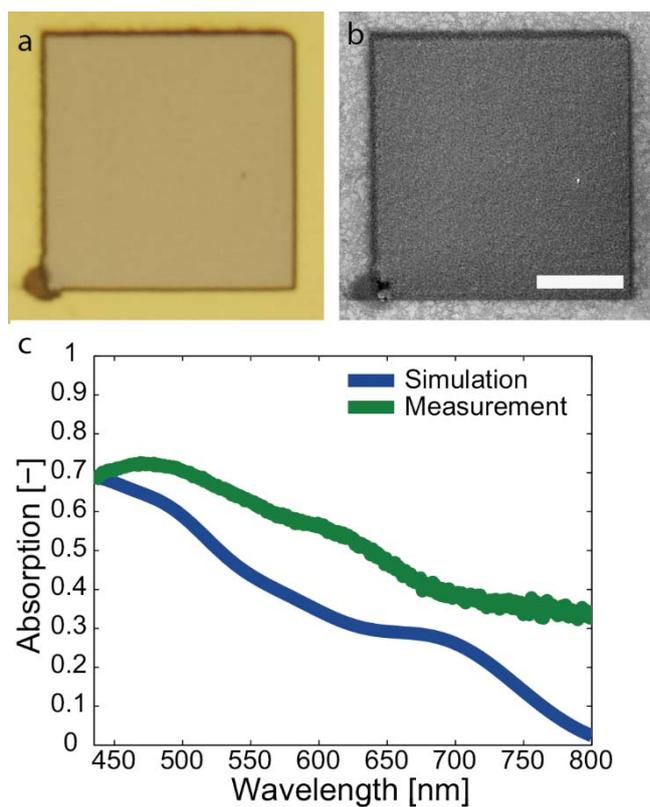

**Figure S3**. Unstructured printed gold layer. Opical (a) and SEM (b) image of unstructured gold nanoparticle pad. The ligand is decanethiol. (c) The absorption of the pad depicted in (a) and (b) and the absorption of a 200 nm thick simulated nanoparticle layer.

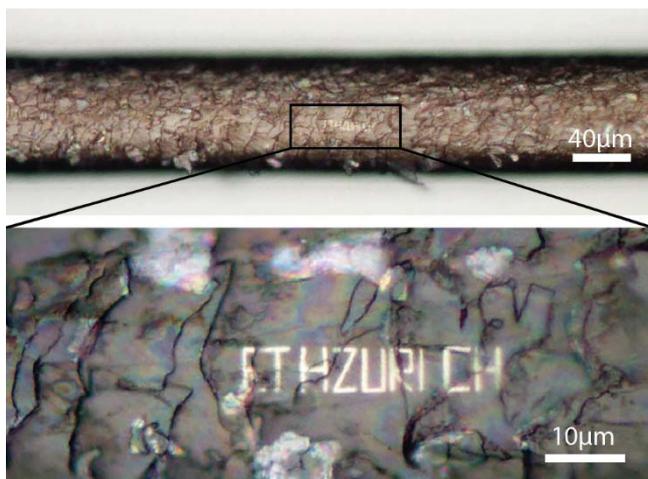

**Figure S4.** Capability to print on soft substrates. Decanethiol-coated gold nanoparticles are printed on an untreated human hair. The three-dimensional, soft and organic structure does not inhibit printing.

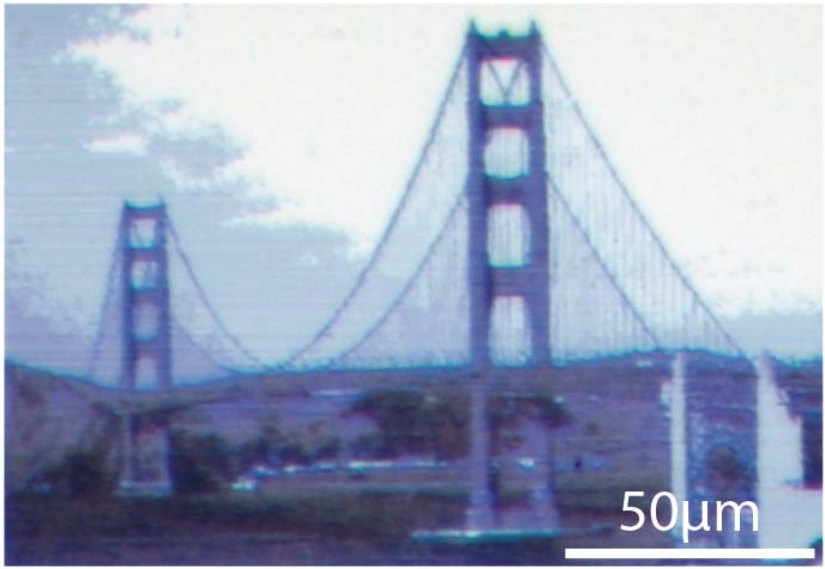

**Figure S5.** Printed picture of the Golden Gate Bridge with a pillar-to-pillar pitch of 400nm. The substrate is optically thick silver[12] with 60nm $Al_2O_3$ as a spacer layer.